\documentclass[iop]{emulateapj}
\usepackage{multirow}
\usepackage{graphicx}
\usepackage{color}

\newcommand{\vlsr}{V$_{\rm LSR}$}
\newcommand{\kms}{km~s$^{-1}$}
\newcommand{\Lsun}{L$_\odot$}
\newcommand{\Msun}{M$_\odot$}

\slugcomment{accepted for publication in ApJ}

\shorttitle{Ionized Envelopes of Orion SrcI and BN}
\shortauthors{Plambeck et al.}

\begin{document}

\title{The Ionized Circumstellar Envelopes of Orion Source I \\ and the Becklin-Neugebauer Object}

\author{R.~L.~Plambeck\altaffilmark{1},
   A.~D.~Bolatto\altaffilmark{2},
   J.~M.~Carpenter\altaffilmark{3},
   J.~A.~Eisner\altaffilmark{4},
   J.~W.~Lamb\altaffilmark{5},
   E.~M.~Leitch\altaffilmark{5,6},
   D.~P.~Marrone\altaffilmark{4},
   S.~J.~Muchovej\altaffilmark{5},
   L.~M.~P\'{e}rez\altaffilmark{3},
   M.~W.~Pound\altaffilmark{2},
   P.~J.~Teuben\altaffilmark{2},
   N.~H.~Volgenau\altaffilmark{5},
   D.~P.~Woody\altaffilmark{5},
   M.~C.~H.~Wright\altaffilmark{1},
   B.~A.~Zauderer\altaffilmark{7}}

\altaffiltext{1}{Radio Astronomy Lab, Hearst Field Annex, University of California, Berkeley, CA 94720 USA}
\altaffiltext{2}{Department of Astronomy, University of Maryland, College Park, MD 20742 USA}
\altaffiltext{3}{Department of Astronomy, California Institute of Technology, Pasadena, CA 91125 USA}
\altaffiltext{4}{Department of Astronomy, University of Arizona, 933 North Cherry Ave., Tucson, AZ 85721 USA} 
\altaffiltext{5}{Owens Valley Radio Observatory, California Institute of Technology, P.O. Box 968, Big Pine, CA 93513 USA}
\altaffiltext{6}{Department of Astronomy and Astrophysics, University of Chicago, 5640 South Ellis Ave., Chicago, IL 60637 USA}
\altaffiltext{7}{Department of Astronomy, Harvard University, Cambridge, MA 02138 USA}

\begin{abstract}

The 229~GHz ($\lambda$1.3mm) radio emission from Orion-KL was mapped with up to
$0.14''$ angular resolution with CARMA, allowing measurements of the flux
densities of Source I (`SrcI') and the Becklin-Neugebauer Object (BN), the 2
most massive stars in this region.  We find integrated flux densities of
$310\pm45$~mJy for SrcI and $240\pm35$~mJy for BN.  SrcI is optically thick
even at 229~GHz.  No trace of the H30$\alpha$ recombination line is seen in its
spectrum, although the $v_2$=1, \hbox{5(5,0)-6(4,3)} transition of H$_2$O,
3450~K above the ground state, is prominent.  SrcI is elongated at position
angle 140$^\circ$, as in 43~GHz images.  These results are most easily
reconciled with models in which the radio emission from SrcI arises via the
H$^-$ free-free opacity in a T~$<4500$~K disk, as considered by
\citet{Reid2007}.  By contrast, the radio spectrum of BN is consistent with
p$^+$/e$^-$ free-free emission from a dense ($n_e \sim 5 \times
10^7$~cm$^{-3}$), but otherwise conventional, hypercompact HII region.  The
source is becoming optically thin at 229~GHz, and the H30$\alpha$ recombination
line, at \vlsr =$23.2\pm0.5$~\kms, is prominent in its spectrum.  A Lyman
continuum flux of $5\times10^{45}$ photons~s$^{-1}$, consistent with that
expected from a B~star, is required to maintain the ionization.  Supplementary
90 GHz observations were made to measure the H41$\alpha$ and H42$\alpha$
recombination lines toward BN.  Published 43 and 86~GHz data suggest that SrcI
brightened with respect to BN over the 15 year period from 1994 to 2009.

\end{abstract}


\section{Introduction}

The Kleinmann-Low Nebula in Orion is well-known as the nearest region of high
mass star formation, 415~pc away \citep{Menten2007,Kim2008}.  It contains at
least 2 massive young stars.  One of these, the Becklin-Neugebauer Object (BN),
has been studied extensively at infrared wavelengths \citep{Scoville1983}; it
is thought to be a B star.  The other object, Source I (hereafter, `SrcI'), is
so heavily obscured by foreground dust that it is not directly visible in the
infrared, although light reflected by the surrounding nebulosity provides a
glimpse of its spectrum \citep{Morino1998,Testi2010}.  SrcI is noteworthy
because it is surrounded by a cluster of SiO masers, one of the few cases in
which SiO masers are associated with a young star.  It also is known to drive a
bipolar outflow into the surrounding molecular cloud \citep{Plambeck2009}.

Remarkably, proper motion measurements show that SrcI and BN are recoiling from
one another at 35-40 \kms\ \citep{Rodriguez2005,Gomez2008,Goddi2011}.  Tracing
the motions backward, \citet{Goddi2011} find that 560 years ago the projected
separation of these two stars in the plane of the sky was just $50 \pm 100$~AU.
An extensive system of `bullets,' bow shocks, and `fingers,' visible in lines
of H$_2$, FeII, and CO \citep{Allen1993,Zapata2011}, also is centered on
Orion-KL.  Proper motion measurements suggest that the fingers were created by
an explosive event 500-1000 years ago \citep{Doi2002,Bally2011}.  Thus, the
currently favored paradigm for Orion-KL \citep{Gomez2008,Bally2011,Goddi2011}
postulates that SrcI and BN were ejected from a multiple system approximately
500 years ago, and that the ejection of the stars unbound the surrounding gas
and circumstellar disks, creating the finger system.

BN's luminosity of $(0.8-2.1)\times10^4$~\Lsun\ \citep{DeBuizer2012} suggests
that it is a 10 to 15~\Msun\ star \citep{Schaller1992}.  Estimates of the mass
of SrcI are conflicting.  SrcI is recoiling at about half the speed of BN in
the rest frame of the Orion Nebula Cluster \citep{Goddi2011}; conservation of
momentum then suggests that its mass is roughly 20~\Msun.  On the other hand,
if the SiO masers near SrcI are in Keplerian rotation about the star, the
inferred central mass is only 7~\Msun\ \citep{Matthews2010}; this can be
interpreted as a lower limit if the SiO-emitting gas is supported in part by
radiation or magnetic pressure.  SrcI could well be a compact binary with a
semimajor axis of $< 10$ AU \citep{Gomez2008}.

At cm and mm wavelengths the continuum spectra of BN and SrcI are believed to
be dominated by free-free emission from ionized circumstellar gas that is at
least partially optically thick up to 100~GHz \citep{Plambeck1995}.
Observations at higher frequencies, where this emission should become optically
thin, can be used to estimate the Lyman continuum fluxes from the stars, better
constraining their masses.  Such observations must be made with subarcsecond
angular resolution in order to distinguish the compact circumstellar emission
from bright but extended dust and molecular line emission from the surrounding
molecular cloud.

Here we describe high angular resolution 229~GHz observations of Orion-KL with
the Combined Array for Research in Millimeter-Wave Astronomy (CARMA) that
cleanly resolve the emission from both stars.  Surprisingly, we find that the
emission from SrcI is optically thick even at 229~GHz, and we fail to find the
H30$\alpha$ recombination line in its spectrum.  We argue that these results
favor models in which the radio emission from SrcI originates from the H$^-$
opacity in a T $<4500$~K disk.

\section{Observations}

The spectra of SrcI and BN presented here were obtained from a single night's
observation of Orion made with the CARMA B-array on 2011~January~06.  Fourteen
antennas were used; projected antenna separations ranged from 50 to
700~k$\lambda$.  The phase center was 05$^{\rm h}$35$^{\rm m}$14\fs505,
$-05$\degr22\arcmin30\farcs45 (J2000).  The weather was excellent -- the
atmospheric opacity at 225 GHz was 0.10, and the rms atmospheric phase
fluctuations were 65 $\mu$m on a 100-m baseline.  Double sideband system
temperatures for the dual-polarization receivers were 100--200 K, scaled to
outside the atmosphere.  For each polarization the correlator was configured to
observe three overlapping 500~MHz windows (covering the frequency range
225.8--227.2~GHz in the receivers' lower sideband and 231.4--232.8~GHz in the
upper sideband, with 12.5~MHz wide channels), plus a single 250~MHz window
(covering 226.65--226.90~GHz in the lower sideband and 231.75--232.00~GHz in
the upper sideband, with 3.125~MHz resolution).  Upper and lower sideband
signals are separated in the correlator by phase-switching.

The antenna gains were derived from observations of the calibrators 0423-013
(2.4~Jy) and 0607-085 (1.05~Jy) that were interleaved with the Orion scans
every 10 minutes.  To minimize decorrelation from atmospheric phase
fluctuations, a self-calibration interval of 30 seconds was used.  The
calibrator flux densities were established from observations of the primary
flux standard, Uranus.  Because Uranus is heavily resolved on these baselines,
only antenna pairs with separations $< 200$ k$\lambda$ were used for flux
calibration.  Antenna gain solutions were applied to the Orion data after
smoothing to a 10~min interval (scalar averaging the amplitude corrections).
Only data with antenna separations greater than 250~k$\lambda$ were used for
the Orion maps, yielding a $0.28'' \times 0.21''$ synthesized beam at PA
$-27^\circ$.  

With $0.25''$ angular resolution, almost all spectral line emission from the
Orion Hot Core and surrounding molecular cloud is resolved out.  Nevertheless,
to ensure that continuum flux densities were not contaminated by residual line
emission or absorption, we made maps of all the spectral channels, and flagged
those (28 out of 234) with anomalously high ($>2 \times$ average) rms noise;
the noisy channels coincided with strong spectral lines.  We generated a
primary-beam corrected continuum map from the remaining channels, using
multifrequency synthesis to avoid radial smearing.

\begin{figure*} [t]
\begin{center}
\includegraphics[angle=-90,scale=0.7]{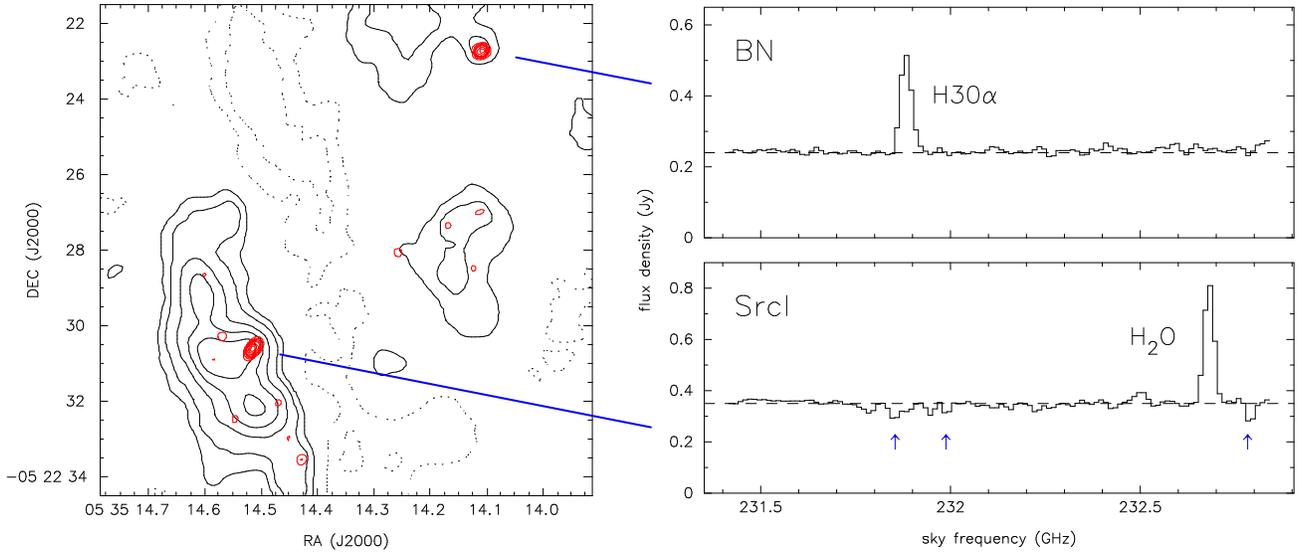}
\caption { \label{fig1} ({\it left}) 229 GHz continuum maps of Orion-KL from
CARMA C- and B-array data.  Black contours show the $0.83''$ resolution C-array
map; the contour levels are $-100, -50$, 50, 100, 200, 300, and 400 mJy/beam.
Negative contours are attributable to poorly sampled extended emission. Red
contours show the $0.25''$ resolution B-array image; contour levels are 20, 40,
60, 80, 100, 150, 200, and 250 mJy/beam.  Extended emission is almost entirely
filtered out in this map; the most negative value is $-30$~mJy/beam.
({\it right})  Spectra of BN and SrcI near 232~GHz obtained from the $0.25''$
resolution image.  Dashed lines indicate the continuum levels.  The spectra
have been Hanning smoothed to 30 \kms\ velocity resolution. The H30$\alpha$
line is detected toward BN, the H$_2$O $v_2$=1, 5(5,0)-6(4,3) line toward SrcI.
Arrows in the SrcI spectrum indicate weak absorption features from transitions
of ethyl cyanide, dimethyl ether, and methanol.}
\end{center}
\end{figure*}

To help correct the Orion data for atmospheric decorrelation, a single
phase-only self-calibration was performed on the source itself.  This means
that after cleaning the continuum map in the normal way, the brightest pixels
in the clean component list were used as a source model to refine the
antenna-based phases.  A time interval of 2 minutes was used, short enough to
track the most significant atmospheric phase variations, but long enough to
obtain a signal to noise ratio of $\geq$1 on each of the 91 available
baselines, allowing a robust least squares fit for the 13 antenna-based phases.
Applying the self-calibrated phase corrections increased the source flux
densities by about 50\%.  The measured rms noise in the final $0.25''$ resolution
map is 4~mJy beam$^{-1}$.

Even higher resolution images of Orion were obtained in 2009 February using the
CARMA A-array, with antenna separations of up to 1.7~km (1350 k$\lambda$).
These observations utilized older, single-polarization receivers and a 1.5~GHz
bandwidth correlator.  For these data the CARMA Paired Antenna Calibration
System \citep[\hbox{``C-PACS,''}][]{Perez2010} was used as an adaptive optics
scheme to correct for blurring by atmospheric phase fluctuations.
\hbox{C-PACS} placed the 8 CARMA 3.5-meter telescopes next to a subset of the
10-m and 6-m telescopes, including those at the extremes of the array.  As the
\hbox{6-m} and 10-m telescopes observed Orion at 229~GHz, the 3.5-m antennas
observed the nearby (1.6$^\circ$ away) quasar 0541-056 at 31~GHz in order to
monitor the atmospheric phase variations above each station.  The atmospheric
phase delay to each 3.5-m telescope was derived by self-calibration on
0541-056 every 12~sec. These antenna-based phases were scaled up by the ratio
of the observing frequencies (229/31 = 7.4) and applied to the mm data.  The
\hbox{C-PACS} calibration accuracy is limited by the angular separation of the
cm calibrator and the mm source, which causes the cm and mm beams to probe
different paths through the atmosphere; by the signal to noise ratio on the
31~GHz calibrator; and by the airmass.

A continuum map was generated from the A-array data using only baselines longer
than 250~k$\lambda$.  The synthesized beam was $0.15'' \times 0.13''$ at PA
14\degr.  Applying the \hbox{C-PACS} corrections more than doubled the signal to noise
ratio of this map.  As with the B-array map, we then performed a single
phase-only self-calibration, with a 2 minute interval, to help correct for
residual atmospheric phase fluctuations; this doubled the flux densities of the
compact sources.  The measured rms noise in the final $0.14''$ resolution map
is 2.6~mJy beam$^{-1}$.

The integrated flux densities measured from the $0.14''$ and $0.25''$
resolution maps were 240 and 255~mJy for BN, 310 and 370~mJy for SrcI.
Although the discrepancy in the two SrcI flux densities could conceivably be
caused by time variability of the source (cf.  section~\ref{timevariable}), it
is more likely attributable to differences in the sampling of visibilities by
the A and B arrays, which lead to different sidelobe structure from the
adjacent Orion Hot Core.  We will use the A-array flux densities for the
remainder of this paper because the higher resolution data should more
effectively filter out emission from the Hot Core.  We estimate that the absolute
flux scale is accurate within $\pm15$\%.

Finally, supplementary 3mm observations were made with the A-array in 2009
February to obtain spectra of the H41$\alpha$ (92.034~GHz) and H42$\alpha$
(85.688~GHz) recombination lines toward BN.  Both of the recombination lines
and the strong SiO maser at 86.243 GHz were measured simultaneously.  The data
were self-calibrated on the maser using a 10 sec interval.  The spectra of the
two recombination lines were averaged together to improve the signal to noise
ratio.

\begin{figure} [b] 
\begin{center}
\vspace{10pt}
\epsscale{1.0}   
\plottwo{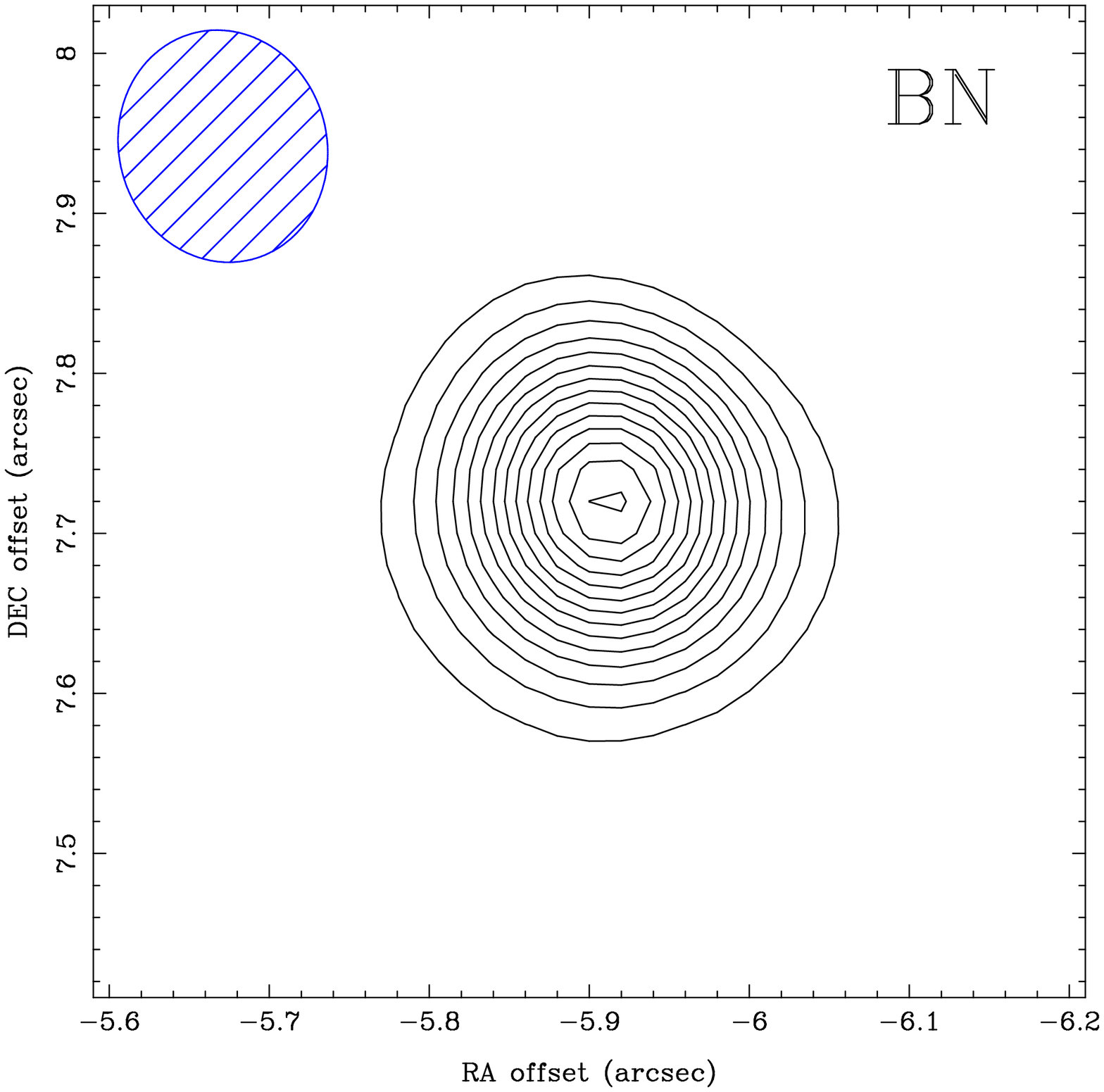}{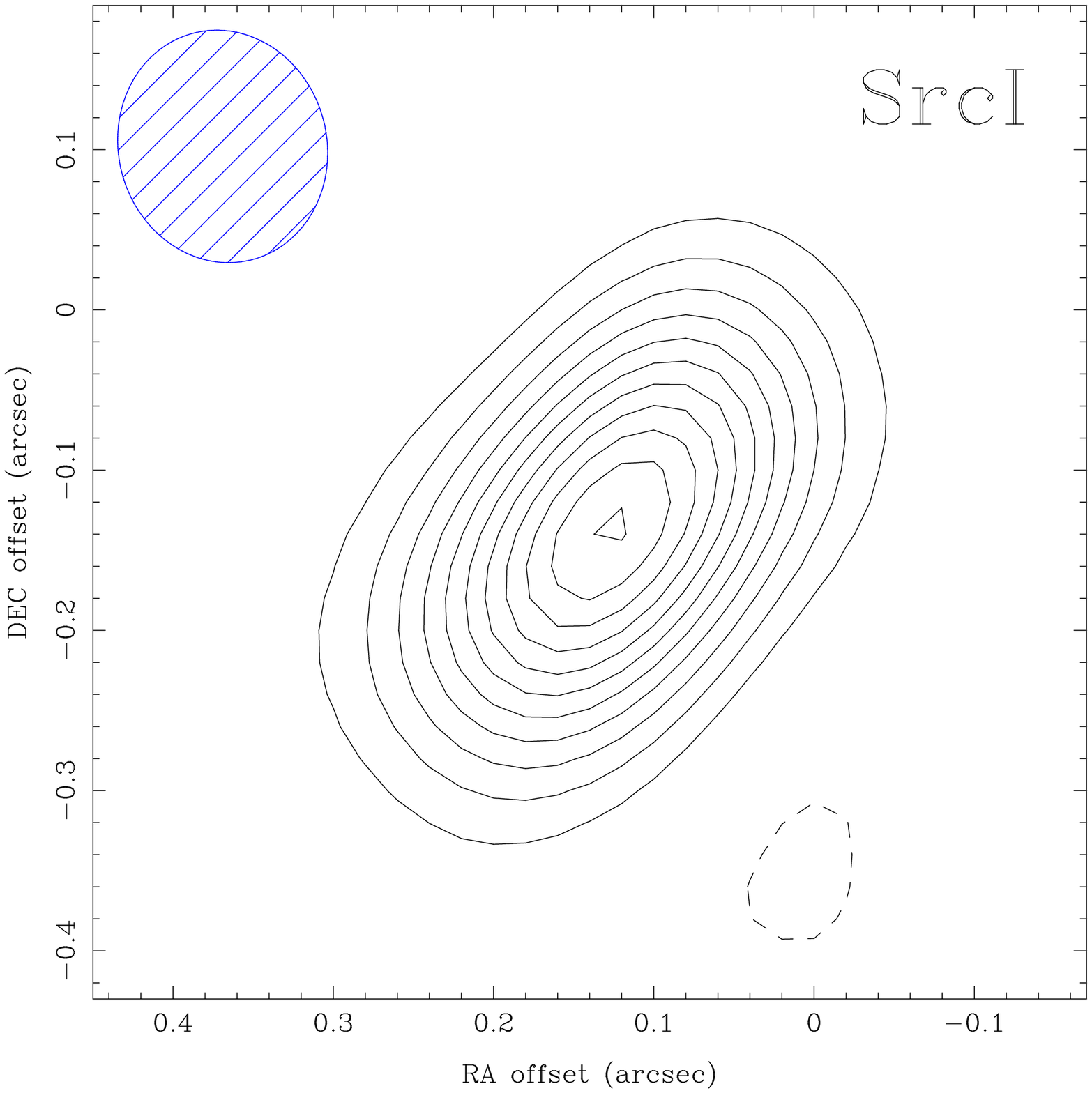} 
\caption{ \label{fighighres} Images of BN and SrcI at 229 GHz from the CARMA
A-array data.  Each box is $0.6''$ on a side.  The lowest contour and the
contour interval is 15~mJy~beam$^{-1}$.  The rms noise level is 2.6 mJy
beam$^{-1}$.   The $0.15'' \times 0.13''$ synthesized beam is shown by a
hatched ellipse in the upper left corner of each image.  SrcI is extended at PA
140\degr, as in 43~GHz VLA images \citep{Reid2007, Goddi2011}.} 
\end{center}
\end{figure}

\section{Results}

The left hand panel in Figure~\ref{fig1} shows the $0.25''$ resolution
continuum map from the B-array observations, overlaid on a $0.83''$ resolution
image obtained with the CARMA C-array \citep{Eisner2008}.  SrcI and BN are
clearly distinguishable only in the higher resolution map; extended emission
from dust and molecular lines dominate the lower resolution image.  

The upper sideband spectra of BN and SrcI are shown in the right hand panels of
Figure~\ref{fig1}.  The H30$\alpha$ recombination line at 231.901~GHz is
prominent in the spectrum of BN, but is not detected toward SrcI.  The bright
line at 232.687~GHz in SrcI's spectrum is the $v_2$=1, 5(5,0)-6(4,3)
transition of H$_2$O.  Weak absorption features toward SrcI are due to
transitions of ethyl cyanide, dimethyl ether, and methanol.  No emission lines
were detected toward either source in the lower sideband.

\begin{deluxetable*}{lccccc}[t]
\tablecaption{ \label{srctable} Source parameters for SrcI and BN at 229 GHz}
\tablewidth{0pt}
\tablehead{
\colhead{source} & \colhead{RA} & \colhead{DEC} & \colhead{peak
flux density} & \colhead{integrated flux density} & \colhead{deconvolved size} \\
& \colhead{( h  m  s )} & \colhead{($^\circ$ $'$ $''$)} & \colhead{(mJy)} & \colhead{(mJy)}
&  \\ }
\startdata
SrcI   &  05 35 14.514  &  -05 22 30.59  &  $170 \pm 25$  &  $310 \pm 45$ &  $0.20'' \times 0.03''$ at PA $140^\circ$ \\
BN     &  05 35 14.109  &  -05 22 22.73  &  $215 \pm 30$  &  $240 \pm 35$ &  $0.06'' \times 0.04''$ at PA $90^\circ$ \\
\enddata
\tablecomments{All parameters were measured from the $0.14''$
resolution map.  Positions are for epoch 2009.1.}
\end{deluxetable*}

Figure~\ref{fighighres} displays the $0.14''$ resolution images of BN and SrcI 
from the A-array.  BN is nearly circularly symmetric, while SrcI is elongated.
Table~\ref{srctable} summarizes the results of Gaussian fits to the positions,
flux densities, and deconvolved sizes.  The 229~GHz source positions match the
43~GHz positions measured in 2009 January by \citet{Goddi2011} within
$0.015''$.  The deconvolved size of SrcI at 229~GHz, $0.20''\times0.03''$ at PA
140$^\circ$, is similar to the sizes measured at 8.4~GHz ($0.19''\times<0.15''$
at PA $136^\circ$; \citealt{Gomez2008}) and at 43~GHz ($0.23''\times0.12''$ at
PA $142^\circ$; \citealt{Goddi2011}), although the source appears to be thinner
along its minor axis at 229~GHz.

\begin{figure*}
\begin{center}
\epsscale{1.} 
\plotone{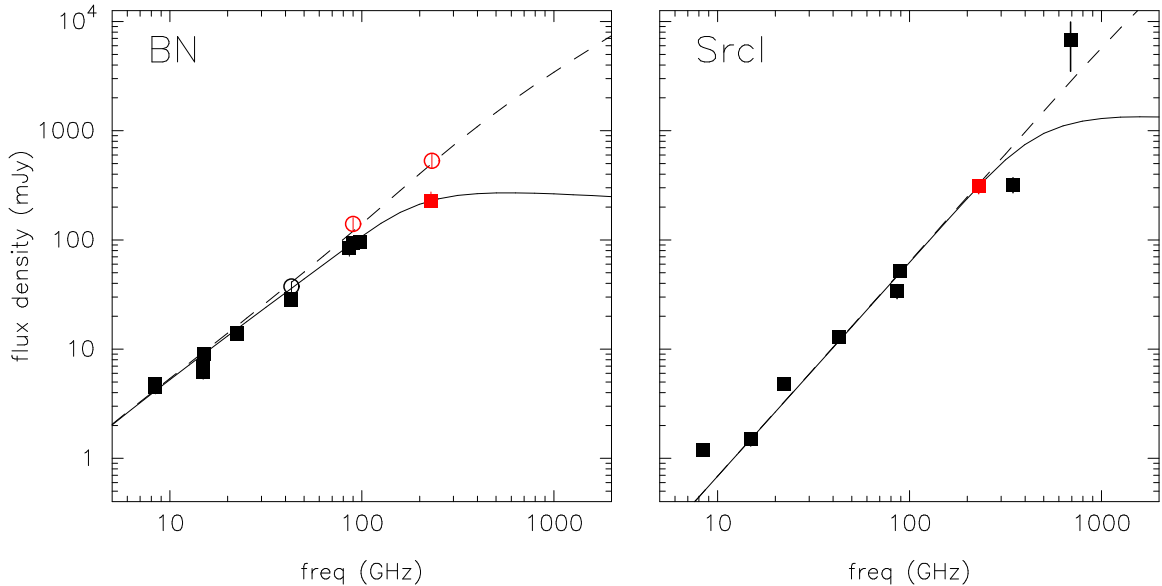}  
\caption{ \label{fluxfig} Radio spectra of BN and SrcI.  Squares indicate
continuum flux densities; circles, flux densities at the peak of recombination
lines.  CARMA results presented in this paper are shown as red in the online
edition.  Solid and dashed lines show, respectively, the continuum and
recombination line flux densities predicted by the models described in sections
4.1.1 and 4.2.1.  }
\end{center}
\end{figure*}

The flux density we measure for BN in the $0.14''$ resolution map is roughly
$2\times$ higher than the value recently reported by \citet{GalvanMadrid2012}
from an analysis of ALMA Band 6 Science Verification data.  These early ALMA
data had a $1.3''\times0.6''$ synthesized beam.   We suspect that the flux
density of BN is depressed in the ALMA map because of negative sidelobes from
poorly sampled extended emission.  The same problem affects the $0.83''$
resolution CARMA image shown in Figure~\ref{fig1}; the peak intensity at the
position of BN in this low resolution image is $\sim125$~mJy/beam.  The higher
resolution CARMA maps are free of deep negative sidelobes, allowing more
reliable flux density measurements of compact sources.

\section{Discussion}

Figure~\ref{fluxfig} shows the radio continuum spectra of BN and SrcI from cm
to submillimeter wavelengths.  Numerical values are given in
Table~\ref{fluxtable}.  For BN we also plot the peak intensities of the
H30$\alpha$ and H41/42$\alpha$ recombination lines measured with CARMA, and the
H53$\alpha$ line measured with the VLA \citep{Rodriguez2009}.  

\subsection{BN}

The continuum spectrum of BN is typical of free-free emission from a
hypercompact HII region. The flux density scales as $\nu^{1.3}$, indicating
that optically thick emission has a larger angular extent at lower frequencies,
either because the ionized gas is clumpy \citep{Ignace2004} or because the
electron density declines smoothly with radius \citep{Wright1975}.  The
spectrum begins to flatten at 229 GHz, indicating that the source is becoming
optically thin at higher frequencies.

\begin{deluxetable}{ccccc}[b]
\tablecaption{ \label{fluxtable} Integrated flux densities for SrcI and BN}
\tablewidth{0pt}
\tablehead{
\colhead{frequency} & \colhead{S(SrcI)} & \colhead{S(BN)} & \colhead{epoch} & \colhead{References} \\
\colhead{(GHz)}     & \colhead{(mJy)}   & \colhead{(mJy)} & \\ 
}
\startdata
4.8   &  \nodata           &  $2.2 \pm 0.8$    & 1990    & 1 \\  
8.4   &  $1.1 \pm 0.2$     &  $4.5 \pm 0.7$    & 1994.3  & 2 \\  
8.4   &  $1.2 \pm 0.1$     &  $4.8 \pm 0.1$    & 2006.4  & 3 \\  
15  &  \nodata           &  $9.1 \pm 0.6$    & 1981.6  & 4 \\  
15  &  \nodata           &  $9.4 \pm 1.0$    & 1983.7  & 4 \\  
15  &  $1.54 \pm 0.18$   &  $6.19 \pm 0.19$  & 1986.3  & 5 \\  
15  &  $1.6 \pm 0.4$     &  $6.8 \pm 1.5$    & 1990    & 1 \\  
22  &  $\cdots$          &  $13.6\pm 1.1$    & 1983.7  & 4 \\  
22  &  $5.7 \pm 0.9$     &  $16.57 \pm 0.7$  & 1991.5  & 6 \\  
43  &  $13 \pm 2$        &  $31 \pm 4.7$     & 1994.3  & 2 \\  
43  &  $10.8 \pm 0.6$    &  $28.0 \pm 0.6$   & 1994.9  & 7 \\  
43	  &  $13$              &  $26.4 \pm 0.7$   & 2000.9  & 3,8 \\  
43    &  $14.5 \pm 0.7$    &  $28.6 \pm 0.6$   & 2007.9  & 9 \\  
43	  &  $11 \pm 2$        &  $23 \pm 2$       & 2009.0  & 10 \\ 
86    &  $34 \pm 5$        &  $84 \pm 10$      & 1995.0  & 11 \\ 
89    &  $50 \pm 5$        &  $91 \pm 9$       & 2009.0  & 12 \\ 
229   &  $310 \pm 45$      &  $240 \pm 35$     & 2009.1  & 13 \\ 
348	  &  $320 \pm 48$      &  \nodata          & 2004.1  & 14 \\ 
690	  &  $6700 \pm 3200$    &  \nodata          & 2005.1  & 15 \\ 
\enddata

\tablerefs{(1)~\citealt{Felli1993b}; (2)~\citealt{Menten1995};
(3)~\citealt{Gomez2008}; (4)~\citealt{Garay1987}; (5)~\citealt{Felli1993a};
(6)~\citealt{Forbrich2008}; (7)~\citealt{Chandler1997}; (8)~\citealt{Reid2007};
(9)~\citealt{Rodriguez2009}; (10)~\citealt{Goddi2011};
(11)~\citealt{Plambeck1995}; (12)~\citealt{Friedel2011} (assuming
10\% absolute calibration accuracy); (13)~this paper; (14)~\citealt{Beuther2004};
(15)~\citealt{Beuther2006} } \end{deluxetable}


\subsubsection{Model spectra}

To model the spectrum, we assume that the emission region is spherically
symmetric, with a uniform electron temperature $T_e$ and an electron density
that is constant, $n_e(r) = n_0$, inside core radius $r_c$, then declines as
$n_e(r) = n_0\,(r/r_c)^{\alpha}$ at larger radii.  For each projected radius
$x$ we numerically integrate along the line of sight to compute the emission
measure $EM(x) = \int \!  n_e^2\,dz$ pc~cm$^{-6}$.  The brightness temperatures
at this projected radius in the continuum and at the peak of a recombination
line are then $T_C(x) = T_e[1-{\rm exp}(-\tau_C)]$ and $T_L(x)=T_e[1-{\rm
exp}(-\tau_C-\tau_L)]$, where the continuum and line opacities are approximated
as \citep{Rohlfs2000}
\begin{eqnarray}
\tau_C(x) & = & 8.23 \times 10^{-2}\,T_e^{-1.35}\,\nu_{\rm GHz}^{-2.1}\,EM(x) \\
\tau_L(x) & = & 1.92 \times 10^3\ T_e^{-2.5}\ EM(x)\,\Delta\nu^{-1}_{\rm kHz}.
\end{eqnarray}
We integrate the brightness temperatures over circular annuli to obtain
flux densities, assuming a distance to the source of 415 pc.

The model BN spectra shown by smooth curves in Figure~\ref{fluxfig} assumed
$T_e = 8000$~K, $n_0=5\times10^7$ cm$^{-3}$, $r_c=7.4$~AU, $\alpha=-3.5$, and
recombination line velocity widths of 30~\kms\ for all transitions.  The model
does not take into account systematic velocity gradients or pressure broadening
of the recombination lines.  The model parameters were determined by trial and
error, and are not a formal fit to the data. To first order, however, the
turnover frequency sets $n_0$, the spectral slope sets $\alpha$, and the
absolute flux densities set $r_c$.

\subsubsection{Excitation parameter}

The model parameters above yield an excitation parameter $U=r_c(n_e n_H)^{1/3}
= 5$~pc~cm$^{-2}$, which serves as a measure of the flux of ionizing photons
from the central star.  This corresponds to the excitation parameter expected
for a main sequence B star with Lyman continuum flux $5\times10^{45}$ photons
s$^{-1}$ and total luminosity $(0.5-1)\times10^4$~\Lsun\ \citep{Panagia1973}.
The ionizing flux is comparable to the value derived by \citet{Scoville1983}
from infrared data, and the luminosity is comparable with the bolometric
luminosity obtained by \citet{DeBuizer2012} from 6--37 $\mu$m SOFIA
observations.

\begin{figure} [b] 
\begin{center}
\vspace{10pt}
\epsscale{1}   
\plotone{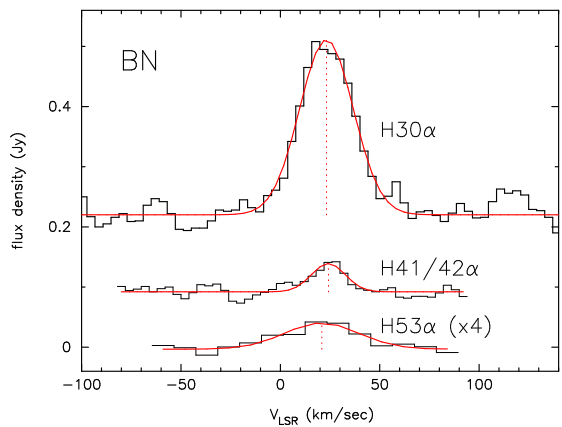}  
\caption{ \label{recombfig} Hydrogen recombination lines observed toward BN.
H30$\alpha$ and H41/42$\alpha$ (the average of the two spectra) were observed
with CARMA; H53$\alpha$, with the VLA \citep{Rodriguez2009}.  The H53$\alpha$
flux densities are scaled up by 4 for better visibility.  Smooth curves
show Gaussian fits to the spectra. } 
\end{center} 
\end{figure}

\subsubsection{Recombination lines}

Figure~\ref{recombfig} displays the recombination line profiles toward BN.
H30$\alpha$ and H41/42$\alpha$ (the average of the H41$\alpha$ and H42$\alpha$
lines) were observed with CARMA, H53$\alpha$ with the VLA
\citep{Rodriguez2009}.  The H30$\alpha$ spectrum was measured using the 250~MHz
wide correlator window with 4~\kms\ channel spacing; it has been Hanning
smoothed to 8~\kms\ resolution.  Line parameters and uncertainties derived from
Gaussian fits to the spectra are summarized in Table~\ref{recombtable}.  The
H53$\alpha$ line is at slightly lower velocity and is noticeably wider than the
higher frequency lines.

For the electron densities $n_e > 10^7$~cm$^{-3}$ that we infer for BN,
collisional broadening of recombination lines can be significant.  Collisional
broadening is most important for high-$n$ transitions because the diameter of a
hydrogen atom is proportional to $n^2$ -- an atom in the n=53 level is 0.3
$\mu$m across!  The collision-broadened linewidths are given by the approximate
formula \citep{Brocklehurst1972} 
\begin{equation} \Delta v_c/2 =
4.7\,c/\nu\,(n/100)^{4.4}\,(10^4/T_e)^{0.1}\, n_e.  
\end{equation} 
For $n_e=5\times10^7$~cm$^{-3}$ and $T_e=8000$~K, $\Delta v_c \sim 200$~\kms\ for
H53$\alpha$, 30~\kms\ for H41$\alpha$, and $3$~\kms\ for H30$\alpha$.  These
widths must be convolved with the expected $\sim20$~\kms\ wide thermal
profiles, plus any additional Doppler broadening from turbulent motions.

Since the observed width of the H53$\alpha$ line is much less than 200~\kms,
most of the H53$\alpha$ emission that is detected must originate more than
$\sim 12$~AU from the star, in the lower density halo where $n_e <
10^7$~cm$^{-3}$.  The line is slightly blueshifted relative to the other
transitions, as expected if this gas is expanding.  The H30$\alpha$ line is
little affected by pressure broadening or by continuum opacity, so its central
velocity, 23.2~\kms, is the best indicator of the \vlsr\ of BN.  We attribute
the anomalously narrow width observed for the H41/42$\alpha$ transition to poor
signal to noise; more sensitive spectra of all these recombination lines would
be useful in further constraining models of the electron density in BN. 

We do not detect the 231.995~GHz helium recombination line toward BN.  The
absence of this line is consistent with the identification of BN as a B-star.
In the 30~\kms\ resolution BN spectrum in Figure~\ref{fig1}, the 2$\sigma$
upper limit on the intensity of the He30$\alpha$ line is $\sim12$~mJy, while
the intensity of the H30$\alpha$ line is 270~mJy; thus
$\tau($He30$\alpha)/\tau($H30$\alpha) < 0.045$.  If helium were ionized over
the full volume of the HII region, the opacity ratio would be comparable to the
He/H abundance ratio, $\sim 0.1$.  Thus we can say that the He$^+$ zone, if it
exists, fills less than half the volume of the HII region.  This places an
upper limit of roughly 0.05 on the ratio of helium-ionizing to
hydrogen-ionizing photons from the star \citep[][Fig. 2]{Mezger1974},
consistent with the ratio expected for B-stars, but not for O9 or more massive
stars \citep{Mezger1974,Vacca1996}.

\begin{deluxetable}{lccccc}[b]
\tablecaption{ \label{recombtable} Recombination Line Parameters for BN}
\tablewidth{0pt}
\tablehead{
\colhead{line} & \colhead{freq} & \colhead{S$_{\rm peak}$\tablenotemark{a}} & \colhead{\vlsr} & \colhead{FWHM} \\
               & \colhead{(GHz)}& \colhead{(mJy)}          & \colhead{(\kms)} & \colhead{(\kms)} \\ 
}
\startdata
H53$\alpha$\tablenotemark{b}  & 42.952   &  $10.4 \pm 1.1$ &  $20.1 \pm 2.1$ & $39.0\pm4.9$ \\
H42$\alpha$  & 85.688   & \multirow{2}{*} {$47\pm5$\tablenotemark{c}} & 
  \multirow{2}{*}{$24.2 \pm 1.1$\tablenotemark{c}}  & 
  \multirow{2}{*}{$20.0 \pm 2.7$\tablenotemark{c}}  \\ 
H41$\alpha$  & 92.034   & & & & \\ 
H30$\alpha$  & 231.901  &  $291 \pm 10$   &  $23.2\pm0.5$  &  $31.8\pm1.3$ \\
\enddata
\tablenotetext{a}{Peak flux density of the line above the continuum level.}
\tablenotetext{b}{\citealt{Rodriguez2009}.}
\tablenotetext{c}{fit to average of the H41$\alpha$ and H42$\alpha$ line profiles.}
\end{deluxetable}

\subsection{SrcI}

According to the data in Figure~\ref{fluxfig}, the flux density of SrcI scales
approximately as $\nu^2$ from 43 to 229 GHz, suggesting that over this
frequency range the source is an optically thick black body with constant
angular size.  Even if the absolute calibrations are wrong, the absence of a
detectable H30$\alpha$ line rules out the possibility that the emission is
optically thin p$^+$/e$^-$ bremsstrahlung, unless the H30$\alpha$ line is
unexpectedly broad ($ > 300$ \kms).  For a linewidth of 30~\kms, comparable to
the velocity span of the SrcI SiO masers, the line to continuum opacity ratio
$\tau_L/\tau_C\sim3$ from equations (1) and (2), and the line should easily
have been detected if $\tau_C<2$.  

Our data cannot easily be reconciled with the 348~GHz flux density of
$320\pm48$~mJy measured with the SMA \citep{Beuther2004}, which falls below
plausible extrapolations from the lower frequency points.  Such a low 348~GHz
flux density would imply that the free-free emission from SrcI is somewhat
optically thin at 229 GHz, in which case we should have detected the
H30$\alpha$ line.  Possibly the 348~GHz SrcI measurement was corrupted by
negative sidelobes from the bright Hot Core clumps 1-2$''$ east of SrcI, or
possibly SrcI is time-variable (cf. Section~\ref{timevariable}). The 690~GHz
flux density measured with the SMA, $6.7\pm3.2$~Jy \citep{Beuther2006}, lies
{\it above} the $\nu^2$ extrapolation from lower frequencies, but this is
easily explained if there is thermal emission from dust within the synthesized
beam.

\subsubsection{An optically thick hypercompact HII region?}

If SrcI is a conventional hypercompact HII region, it must have an
extraordinarily high emission measure, $>3\times10^{11}$~pc~cm$^{-6}$, in order
to remain optically thick at 229~GHz.  We are able reproduce the observed
continuum flux densities with the simple spherically symmetric model described
in section 4.1.1 if we assume that the ionized region has radius $r_c=7.5$~AU,
electron temperature $T_e = 8000$ K, and uniform electron density $n_e=1.5
\times 10^8$~cm$^{-3}$.  The resulting continuum spectrum is shown by the solid
curve in Figure~\ref{fluxfig}; the dashed curve shows the recombination line
intensities predicted by the model.  The HII region is constrained to have a
sharp outer edge -- any halo of lower density plasma would lead to excess
emission at cm wavelengths.  Collisional broadening of the H30$\alpha$ line is
predicted to be of order 10~\kms, not enough to smear out the line.  The
excitation parameter $U\sim10$~pc~cm$^{-2}$ computed from the radius and
electron density corresponds to an ionizing flux of $4 \times 10^{46}$ photons
s$^{-1}$, consistent with a B0 to B1 main sequence star with luminosity $\sim
10^4$~\Lsun\ \citep{Panagia1973,Smith2002}.  Since the continuum emission is
optically thick, this should be interpreted as a lower limit to the ionizing
flux. 

The source diameter of 15~AU used in the model is dictated by the assumed
electron temperature of 8000~K, the measured flux densities, and the distance
to Orion.  A diameter of 15~AU is, however, inconsistent with the observed size
of SrcI, which is roughly $0.2''$, or 80~AU, along its major axis.  This
suggests that the emission region is clumpy or filamentary so that it does not
fill the synthesized beam.  But it is implausible that all these clumps or
filaments are perfectly sharp-edged, such that every one of them is optically
thick even at 229~GHz.  Generally, one expects a distribution of clumps to
produce emission with a shallower spectral slope \citep{Ignace2004}.

One alternative is that the electron temperature is less than 8000~K.  The
highest resolution measurements of SrcI that are available, 43~GHz VLA
observations with a 34~mas synthesized beam, imply that the source has a
brightness temperature of only 1500~K \citep{Reid2007}. This is much cooler
than is possible in an HII region, however; hydrogen is almost entirely neutral
in dense gas at T$< 4000$~K \citep{Reid1997}.

\subsubsection{H$^-$ free-free emission from a massive disk?}

The absence of hydrogen recombination lines, 1500~K apparent brightness
temperature, and spectral index of $\sim$2 all are difficult to explain if SrcI
is a hypercompact HII region.   The data more naturally fit an alternative
hypothesis, first considered by \citet{Reid2007}, that the emission from SrcI
originates in a T$<4500$~K disk, similar to the radio photospheres of Mira
variables \citep{Reid1997}.  The infrared spectrum of SrcI, seen via light
reflected off the neighboring nebulosity, suggests such an interpretation as
well -- \citet{Testi2010} find that the spectrum is characteristic of a
low-gravity photosphere with effective temperature 3500--4500~K.   The close
association of SrcI with SiO masers, also characteristic of Mira variables,
further strengthens the case for this hypothesis.

As discussed by \citet{Reid1997}, free-free emission at temperatures of about
1500~K in such a photosphere arises via the H$^-$ opacity, the scattering of
electrons by neutral hydrogen atoms or molecules.  The electrons are produced
by collisional ionization of Na, K, and other metals.  The H$^-$ opacity is
roughly $10^3$ times smaller than the p$^+$/e$^-$ opacity, so high densities
($>10^{11}$ cm$^{-3}$) typically are required to obtain optically thick
emission.

\citet{Reid2007} were able to fit the 43~GHz VLA observations of SrcI with
models of H$^-$ emission from an 80~AU diameter, 3~\Msun\ disk.  However, they
found that such a large disk cannot be maintained at T~$\geq1500$~K if it is
heated solely by a central star -- a stellar luminosity $>10^5$~\Lsun, greater
than the luminosity of the entire Orion-KL region, would be required.  They
proposed that accretion processes might heat the disk locally.
\citet{Testi2010} also find that the infrared spectrum of SrcI can be explained
by emission from a disk around a $\sim10$~\Msun\ protostar that is accreting at
a few~$\times\ 10^{-3}$~\Msun\ yr$^{-1}$. However, their models show that the
disk temperature falls below 1500~K for r$> 10$~AU, so cannot easily explain the
H$^-$ free-free emission at larger radii.

In summary, the H$^-$ hypothesis explains the observational results in a
natural way. The principal barrier to its acceptance is uncertainty over the
mechanism by which such a large disk can be maintained at sufficiently high
temperature.  Whether the disk could be heated by an anomalously high accretion
rate, a merging binary, magnetic effects, or some other mechanism remains a key
question.

\begin{figure}  
\begin{center}
\epsscale{1.1}   
\plotone{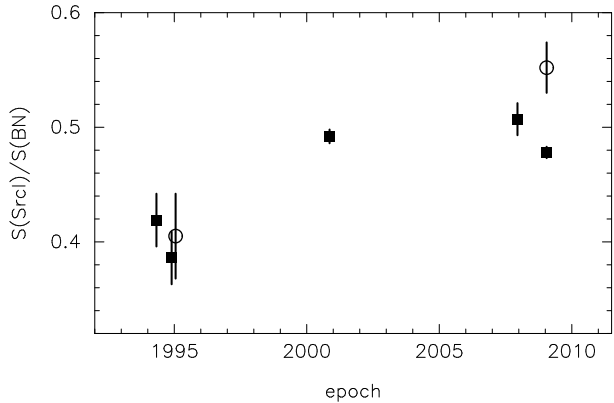} 
\caption{ \label{ratiofig}
SrcI/BN flux density ratio vs. year, at 43 GHz (filled squares) and at
86 GHz (circles), derived from the data in Table~\ref{fluxtable}.  This
ratio should be immune to errors in the absolute flux scale,
since the SrcI and BN are just $10''$ apart on the sky and are observed
simultaneously.  These data provide possible evidence that SrcI has brightened
relative to BN since 1995.}
\end{center} 
\end{figure}

\subsubsection{\label{timevariable} Is SrcI brightening?}

If accretion is responsible for much of SrcI's luminosity, then one might
expect to see changes in its flux density on timescales of years.  Although
\citet{Goddi2011} found no evidence of flux variations at~43 GHz over 4 epochs
from 2000 to 2009, we are struck by the discrepancy in the 86~GHz fluxes
measured at BIMA in 1995 \citep[$34 \pm 5$ mJy,][]{Plambeck1995} and at CARMA
in 2009 \citep[$50 \pm 5$ mJy,][]{Friedel2011}. 

Generally, uncertainties in {\it absolute} flux densities are dominated by
systematic effects -- e.g., variations in telescope gains as a function of
elevation, or uncertainties in the brightness temperature models of planets
used as primary flux standards.  The {\it ratio} of the flux densities of SrcI
and BN should be immune to such effects because the two sources are just $10''$
apart and can be observed simultaneously.  In Figure~\ref{ratiofig} we plot
this ratio as a function of time for the 43 and 86~GHz data in
Table~\ref{fluxtable}.  Note that the error bars in this figure were calculated
from the thermal noise in the original maps, not from the absolute flux
uncertainties listed in Table~\ref{fluxtable}. From these 7 independent
datasets it appears that the ratio increased from about 0.4 in 1995 to 0.5 in
2009.  

This result should be interpreted cautiously.  Although the flux density ratio
is unaffected by the absolute calibration scale, it is sensitive to the
sampling of visibility data in the $(u,v)$ plane.  It is also unclear whether
SrcI or BN, or both, are varying; periodic variations in the infrared
luminosity of BN were reported by \citet{Hillenbrand2001}.  Future careful
monitoring of the flux densities of both SrcI and BN with the VLA and ALMA will
be required to find conclusive evidence for source brightness variations.

\subsubsection{Vibrationally excited H$_2$O}

The bright emission line in the spectrum of SrcI in Figure 1 is the H$_2$O
$v_2$=1, 5(5,0)-6(4,3) transition of H$_2$O at 232.687~GHz, 3450~K above the
ground state.  This line also was recognized by \citet{Hirota2012} in the Orion
Band 6 ALMA Science Verification data.  With 0.6~\kms\ velocity resolution, the
ALMA data show that the H$_2$O line is double-peaked, like the 22 GHz H$_2$O
masers or the 43 and 86 GHz SiO masers toward SrcI.  

In the $1.5''$ resolution ALMA maps the H$_2$O line is blended with a
HCOOCH$_3$ transition at 232.684 GHz, whereas in the $0.25''$ resolution CARMA
map the HCOOCH$_3$ emission is resolved out.  The H$_2$O emission is unresolved
by the CARMA beam, so it is clear that the line originates from the compact SiO
maser zone close to SrcI, rather than the extended $2'' \times 0.5''$ strip
where 22 GHz H$_2$O masers are found \citep{Gaume1998}.  The SiO v=2 masers
also originate in energy levels 3500~K above the ground state;
\citet{Goddi2009} find that these masers can be excited in gas with kinetic
temperature $> 2000$~K.

The same transition of vibrationally excited H$_2$O was detected toward the
evolved stars VY~CMa and W~Hya by \citet{Menten1989}.  Like SrcI, both of these
stars have SiO and 22~GHz H$_2$O masers.

\section{Summary}

CARMA 229 GHz maps of Orion-KL with up to $0.14''$ angular resolution clearly
distinguish the ionized envelopes of SrcI and BN from the surrounding molecular
cloud.  The principal results are as follows:

\begin{enumerate}

\item The integrated flux densities of SrcI and BN at 229 GHz are 310 and
240~mJy respectively, with a probable uncertainty of $\pm$15\%.  The BN flux
density we measure with $0.14''$ resolution is a factor of two greater than
that derived by \citet{GalvanMadrid2012} from ALMA Science Verification data
with a $1.3'' \times 0.6''$ synthesized beam.  We suspect that the flux density
of BN is depressed in the ALMA map because of negative sidelobes from poorly
sampled extended emission.  These large scale structures are filtered out more
effectively in the higher resolution CARMA data.

\item SrcI's flux density is proportional to $\nu^2$ from 43 to 229~GHz -- it
appears to be an optically thick black body.  By comparison, BN's flux density
scales as $\nu^{1.3}$ up to about 100~GHz, then flattens; its free-free
continuum is becoming optically thin at 229~GHz.

\item SrcI is elongated at PA 140 degrees in the CARMA images.  Its deconvolved
size is similar at 229 and 43~GHz -- further evidence that it is optically
thick.

\item The H30$\alpha$ recombination line is not detected toward SrcI, though it
is bright toward BN.  The absence of this recombination line is a third piece of
evidence that SrcI is optically thick, or that it is not an HII region at all.

\item The $v_2$=1, 5(5,0)-6(4,3) transition of H$_2$O at 232.687~GHz, 3450~K
above the ground state, is prominent toward SrcI.  The line emission originates
close to the central star, not in the larger zone where 22 GHz H$_2$O masers
are found.

\item While our observational results do not rule out the possibility that SrcI
is an exceptionally dense hypercompact HII region, they are more easily
explained by models of free-free emission via the H$^-$ opacity in a T$<4500$~K
disk, as modeled by \citet{Reid2007}.   

\item A model of a spherically symmetric hypercompact HII region with a diameter
of about 15~AU and electron density $n_e\sim5\times10^7$~cm$^{-3}$ matches both the
continuum and the recombination line intensities toward BN.

\item The excitation parameter $U\sim5$~pc~cm$^{-2}$ derived for BN implies
that its Lyman continuum flux is $5\times10^{45}$ photons s$^{-1}$, consistent
with that expected from a zero age main sequence B star with a total luminosity
of $(0.5-1)\times 10^4$~\Lsun.

\item The H53$\alpha$ recombination line in BN is predicted to have a
collisionally broadened linewidth of 200~\kms.  Presumably the H53$\alpha$
emission observed by \citet{Rodriguez2009}, with a linewidth of about 40~\kms,
originates primarily from lower density ionized gas farther from the central
star.

\item The \vlsr\ of BN is $23.2\pm0.5$~\kms, based on the central velocity of the
231.9~GHz H30$\alpha$ line, which is little affected by pressure broadening or
continuum opacity.
 
\item Published flux densities at both 43 and 86~GHz provide tentative evidence
that SrcI brightened with respect to BN between 1994 and 2009. 

\end{enumerate}

Future ALMA observations of Orion-KL at submillimeter wavelengths will provide
a definitive test of the emission mechanism in SrcI.  If SrcI is an HII region,
it must become optically thin at sufficiently high frequencies.  If the source
were still optically thick at 650~GHz, for example, an O-star with a Lyman
continuum flux $> 10^{49}$~photons s$^{-1}$ would be required to maintain its
ionization, but the luminosity of such a star, $> 10^5$~\Lsun\ 
\citep{Panagia1973}, would exceed the measured luminosity of the entire
Orion-KL nebula.  Then, at frequencies where the continuum is optically thin,
hydrogen recombination lines {\it must} be detectable if hydrogen is ionized.
If these recombination lines are detected, the free-free hypothesis is
confirmed.  If not, or if the continuum flux density continues to increase
following a $\nu^2$ law, the emission must be attributable to the H$^-$
mechanism.

Submillimeter observations of BN, as well as more sensitive observations of its
cm and mm wavelength recombination lines, will be useful in constraining models
of the hypercompact HII region around this star, and in tracking small changes
in the SrcI/BN flux density ratio that could be an indication of time variable
accretion onto these objects.

{\acknowledgements}

Support for CARMA construction was derived from the states of California,
Illinois, and Maryland, the James S. McDonnell Foundation, the Gordon and Betty
Moore Foundation, the Kenneth T. and Eileen L. Norris Foundation, the
University of Chicago, the Associates of the California Institute of
Technology, and the National Science Foundation. Ongoing CARMA development and
operations are supported by the National Science Foundation under a cooperative
agreement, and by the CARMA partner universities. 

{\it Facilities:} \facility{CARMA}.

\end{document}